\documentclass{ws-mpla}

\newcommand{\beq}{\begin{equation}}
\newcommand{\eeq}{\end{equation}}
\newcommand{\bea}{\begin{eqnarray}}
\newcommand{\eea}{\end{eqnarray}}
\newcommand{\bef}{\begin{figure}}
\newcommand{\eef}{\end{figure}}
\newcommand{\bce}{\begin{center}}
\newcommand{\ece}{\end{center}}
\newcommand{\eg}{{\it e.g.}}
\newcommand{\ie}{{\it i.e.}}
\newcommand{\etal}{{\it et al.}}
\def\lsim{\mathrel{\rlap{\lower4pt\hbox{\hskip1pt$\sim$}}
    \raise1pt\hbox{$<$}}}         %less than or approx. symbol
\def\gsim{\mathrel{\rlap{\lower4pt\hbox{\hskip1pt$\sim$}}
    \raise1pt\hbox{$>$}}}         %greater than or approx. symbol

\begin{document}

\markboth{Ralf Rapp}
{Thermal Photons in Strong Interactions}

\catchline{}{}{}{}{}
%
%%%%%%%%%%%%%%%%%%%%%%%%%%%%%%%%%%%%%%%%%%%%%%%%%%%%%%%%%%%%%%%%%%%%

\title{THERMAL PHOTONS IN STRONG INTERACTIONS}

\author{\footnotesize RALF RAPP}

\address{Cyclotron Institute and Department of Physics, Texas A\& M University\\
College Station, Texas 77843-3366, U.S.A. \\
email:rapp@comp.tamu.edu}

\maketitle

\pub{Received (Day Month Year)}{Revised (Day Month Year)}

\begin{abstract}
A brief survey is given on the current status of evaluating thermal
production of photons from a strongly interacting medium. Emphasis is 
put on recent progress in assessing equilibrium emission rates in both
hadronic and quark-gluon matter. We also give an update on the status 
of comparing theoretical calculations with experimental data from 
heavy-ion collisions at the SPS, as well as prospects for RHIC.
Finally, applications of photon rate calculations to 
colorsuperconducting quark matter are discussed.   

\end{abstract}

\ccode{PACS Nos.: include PACS Nos.}

\section{Introduction}	
Electromagnetic probes have a long and extremely successful history as 
valuable agents of the structure of strongly interacting objects. 
Many collective phenomena in atomic nuclei have been discovered through 
gamma ray spectroscopy, and, at much higher energies, deep inelastic 
scattering of electrons has lead to establishing quarks as the 
elementary building blocks of the nucleon, and is still being intensely 
used at Jefferson Laboratory to explore  its nonperturbative structure.
Furthermore, the emission of photons and dileptons has provided key 
insights into the dynamical environments created in heavy-ion 
collisions (see, \eg,  Refs.~\cite{Fein76,Shu78} for early works and 
Refs.~\cite{RW00,Alam01,PT02,GH03,cyr03} for extensive recent reviews). 
Here one hopes to infer properties of transient stages of highly 
excited matter such as its temperature or in-medium modifications 
of its underlying degrees of freedom. 
In this Brief Review we will focus on the role of thermal photons in 
probing hot and dense strongly interacting matter, with emphasis 
on recent developments rather than a complete account as has been 
given elsewhere\cite{PT02}. 

Emission rate 
calculations provide key input for the identification of thermal 
radiation from both compact stars and high-energy heavy-ion collisions, 
and thus for the search of new states of matter in these systems.
Whereas for stars, due to their macroscopic dimension, essentially all
emitted photons are of a (quasi-) thermal origin, this is not the
case for heavy-ion collisions.   
Corresponding transverse-momentum ($q_t$) spectra of {\em direct} 
photons, \ie, those that are not due to final-state hadron decays, 
can be roughly decomposed into the following regimes: 
at sufficiently high $q_t$, {\em prompt} photons from initial 
nucleon-nucleon ($N$-$N$) collisions prevail, due to a characteristic 
power-law dependence on $q_t$, calculable in perturbative Quantum 
Chromodynamics (pQCD). Towards lower $q_t$, radiation from a hot and 
dense medium is expected to take over. The high-$q_t$ part of the 
thermal spectrum will be most sensitive to the hottest phases of the 
produced medium, whereas at low $q_t$ hadron gas (HG) emission is 
likely to dominate\footnote{There is also a contribution 
from a "pre-equilibrium" phase during which re-interactions have not 
yet thermalized the matter.}. To unfold the various components 
in the spectra a reliable calculation of the thermal emission rates 
from both the QGP and HG is mandatory, which furthermore 
need to be convoluted over the space-time evolution of the 
matter (affecting both yield and slope of the thermal components 
in different ways).  
The emission rates are also of interest in their own right, as 
they, in principle, encode information on the elementary excitations
and their in-medium properties in the respective phase\footnote{Even 
though dilepton invariant-mass spectra allow a more 
direct access to in-medium vector-spectral densities, it should be
emphasized that photons and dileptons are intimately related.}. 
%In particular, we will highlight recent progress in 
%the evaluation of rates from both hadronic and QGP phases. 

The article is organized as follows. In Sect.~\ref{sec_rates} we 
present the current status of evaluating thermal emission rates from 
both hadronic and QGP phases, highlighting recent achievements. 
The latter include a complete leading-order result 
%as well as chemical off-equilibrium effects 
for the QGP (Sect.~\ref{ssec_qgp}), and new production 
channels within flavor-$SU(3)$ in the HG (Sect.~\ref{ssec_hg}).
In Sect.~\ref{sec_hic} we turn to applications of model calculations 
to ultrarelativistic heavy-ion collisions (URHICs). We first give a 
brief account of nonthermal sources (Sect.~\ref{ssec_nontherm}), 
as well as of hydrodynamical simulations which are the natural 
space-time framework over which to convolute thermal rates. This  
is followed by comparisons of up-to-date calculations to (very) recent
data from the Super-Proton-Synchrotron (SPS) at CERN and the Relativistic
Heavy-Ion Collider (RHIC) at BNL, as well as possible interpretations 
thereof, and future prospects (Sec.~\ref{ssec_comp}).
In Sect.~\ref{sec_csc} we discuss evaluations of thermal photon
emissivities from colorsuperconducting quark matter
at high densities and low temperatures, which could have some
bearing on astrophysical contexts of supernovae or compact stars.
We finish with concluding remarks in Sect.~\ref{sec_concl}. 

%%%%%%%%%%%%%%%%%%%%%%%%%%%%%%%%%%%%%%%%%
\section{Thermal Photon Emission Rates}
\label{sec_rates}
%%%%%%%%%%%%%%%%%%%%%%%%%%%%%%%%%%%%%%%%%
Commonly employed formalisms to calculate electromagnetic (e.m.)
emission spectra from a thermal medium are finite-temperature
field theory and kinetic theory. To leading order (LO) in the
e.m. coupling,  
$\alpha_{\rm em}$, the field-theoretic expression for
the emission rate per unit 4-volume, $R$, can be cast into the form 
\beq
q_0\frac{dR_\gamma}{d^3q} = -\frac{\alpha_{\rm em}}{\pi^2} \ 
       f^B(q_0;T) \  {\rm Im}\Pi_{\rm em}^T(q_0=q;T) \ . 
\label{corr}
\eeq
It is exact in the strong interactions which are encoded in the e.m. 
current-current correlation function (or photon selfenergy), $\Pi_{em}$ 
(for real photons, \ie, at invariant mass $M^2$=$q_0^2-q^2$=0, only 
transverse polarizations contribute). In kinetic
theory, the rate for a process of type 1+2$\to$3+$\gamma$ reads
\beq
q_0 \frac{dR_\gamma}{d^3q} =\int (\prod_{i=1}^3 \frac{d^3p_i}{2(2\pi)^3E_i})
(2\pi)^4 \delta^{(4)}(p_1+p_2-p_3-q)
\left|{\cal M}\right|^2\frac{f(E_1)f(E_2)[1\pm f(E_3)]}{2(2\pi)^3}
\label{kin}
\eeq
which is convenient if pertinent scattering amplitudes, $\cal M$,
are evaluated in a perturbative expansion.
Nonperturbative (model) calculations at low and intermediate energies,
on the other hand, are more amenable to the correlator formulation, 
Eq.(\ref{corr}). In the hadronic medium, \eg, $\Pi_{em}$ can be directly 
related to vector-meson spectral functions within the vector dominance 
model (VDM).

%%%%%%%%%%%%%%%%%%%%%%%%%%%%%%%%%%%%%%%%%
\subsection{Quark-Gluon Plasma}
\label{ssec_qgp}
\begin{figure}[th]
\centerline{   
\hspace{-0.4cm} \psfig{file=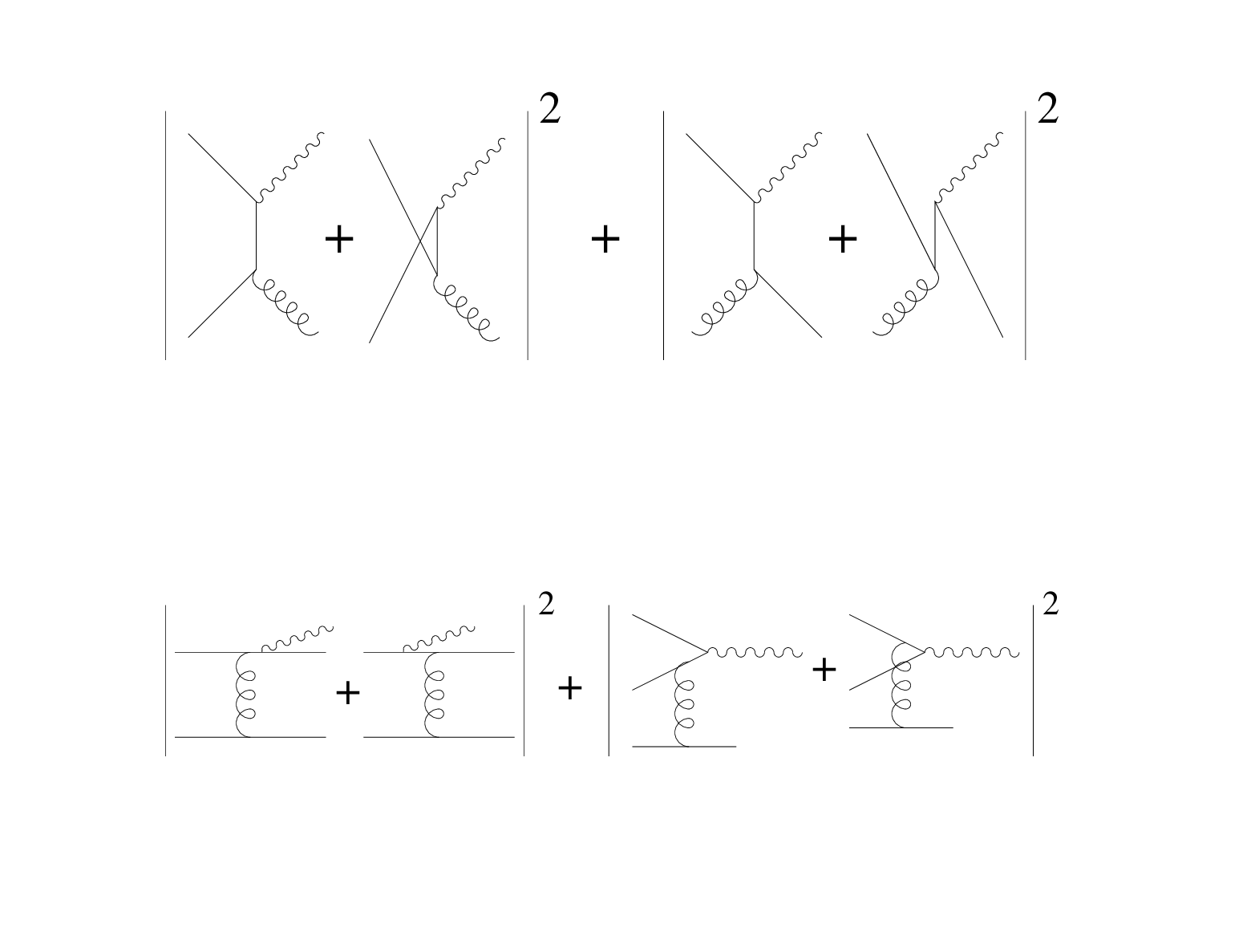,width=7.3cm}
\hspace{-0.4cm} \psfig{file=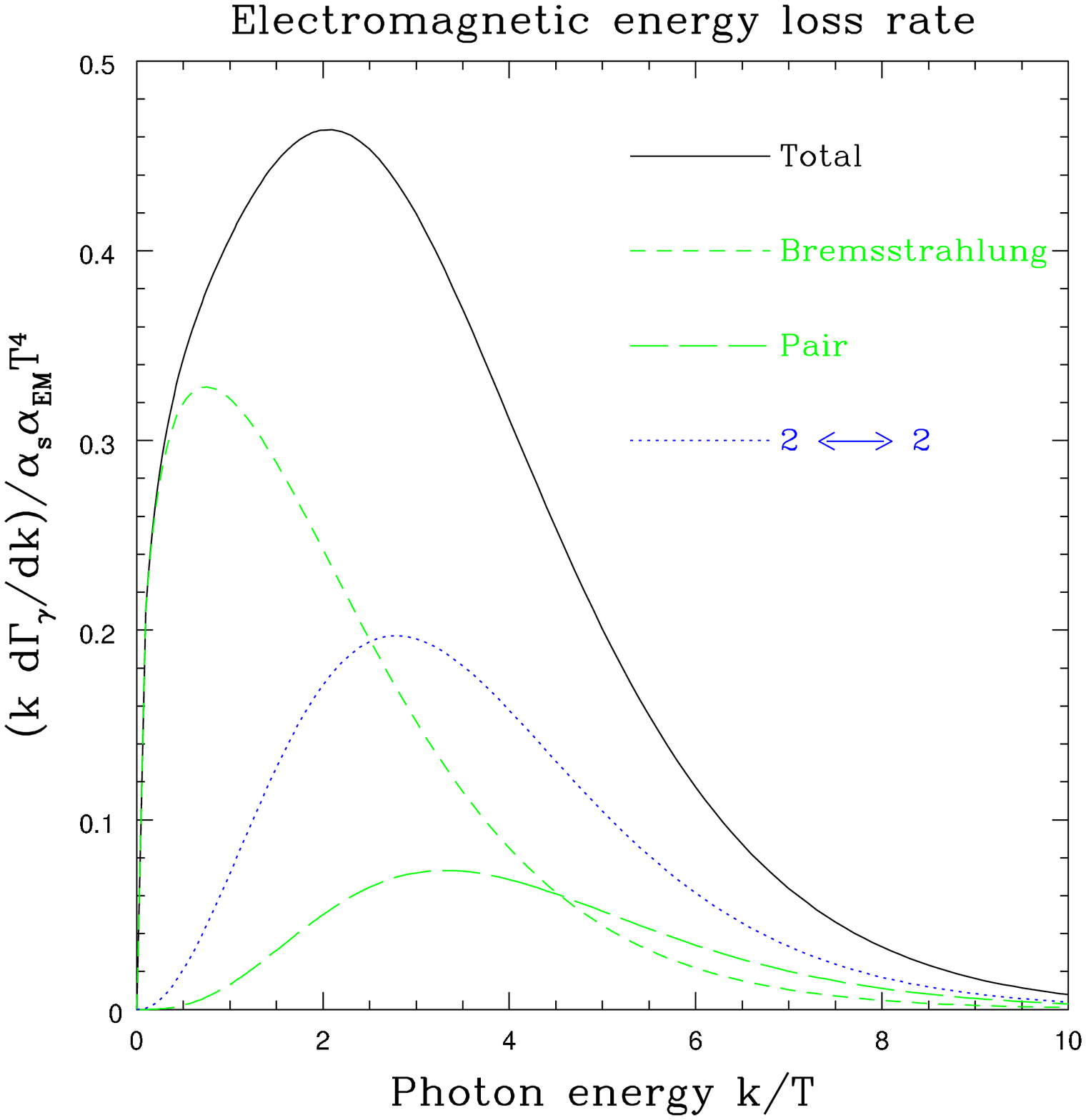,width=5.7cm,height=5.5cm} }
\vspace*{8pt}
\caption{Left panel: leading-order diagrams for photon production 
from a QGP, taken from Ref.~\protect\cite{AMY01} (upper panel:
2$\to$2 scattering; lower panel: Bremsstrahlung and pair
annihilation); right panel: pertinent 
contributions to the photon production rate (multiplied by 
a phase space factor 4$\pi q^2$) in a complete LO 
calculation with $\alpha_s$=0.2\protect\cite{AMY01}.
}
\vspace{-0.2cm}
\label{fig_qgplo}
\end{figure}

%%%%%%%%%%%%%%%%%%%%%%%%%%%%%%%%%%%%%%%%%
First perturbative calculations of the photon production rate from an 
equilibrated QGP at zero net baryon density (\ie, at quark chemical
potential $\mu_q$=0) have been performed
by several groups starting more than 20 years ago\cite{Shu78,KM81}.
It was soon realized\cite{KR83} that forward infrared 
singularities require resummed thermal 
propagators for the exchanged partons. 
For the simplest scattering diagrams depicted in the upper left panel 
of Fig.~\ref{fig_qgplo}, the pertinent result, for large values of  
$x=q_0/T$ and two massless quark flavors ($u$ and $d$), can be
written as\cite{KLS91,Baier92}
\beq
q_0 \frac{dR_\gamma}{d^3q} = \frac{5}{9} \frac{\alpha\alpha_S}{2\pi^2}
       T^2 {\rm e}^{-x}
       \ln\left(1+\frac{2.912}{4\pi\alpha_s} x \right) \  
\label{rate_pert}
\eeq
(an additive ``1" has been introduced in the argument of the logarithm 
to enable extrapolation to small $x$\cite{KLS91}).
As noticed in Ref.~\cite{Aur98} (see also Ref.~\cite{ST01}), 
Eq.~(\ref{rate_pert}) does not yet comprise the full 
result to leading order in the strong coupling constant $\alpha_s$.
Due to collinear singularities, Bremsstrahlung as well as
pair annihilation graphs (cf. lower left panel of Fig.~\ref{fig_qgplo})
contribute at the same order as the resummed 2$\to$2 processes.
The full result, which also necessitates the incorporation of
Landau-Pomeranchuk-Migdal (LPM) interference effects, has been 
computed in Ref.~\cite{AMY01} as  
\beq
q_0 \frac{dR_\gamma}{d^3q} = \frac{5}{9} \frac{\alpha\alpha_S}{2\pi^2}
       T^2 f^F(x) 
      \left[ \ln(\frac{\sqrt{3}}{g}) + \frac{1}{2} \ln(2x) 
      + C_{22}(x) + C_{brems}(x) + C_{ann}(x) \right]    , 
\label{rate_lo}
\eeq
with convenient parameterizations of the 3 functions $C$ given
by\cite{AMY01}
\bea
C_{22}(x)&=&\frac{0.041}{x}-0.3615+1.01{\rm e}^{-1.35x} 
\\
C_{brems}(x) + C_{ann}(x)&=&0.633 x^{-1.5} \ln \left( 12.28+1/x \right)
       +\frac{0.154x}{(1+x/16.27)^{0.5}} \ .  
\eea
The right panel of Fig.~\ref{fig_qgplo} shows the individual 
contributions to the photon production rate corresponding to
 Eq.~(\ref{rate_lo}) (multiplied by an extra phase space factor
4$\pi q^2$). In the for phenomenological applications interesting
range ($x$$\gsim$4) the Bremsstrahlung and pair annihilation 
processes augment the rate from 2$\to$2 scattering by about a 
factor of 2 (see also right panel of Fig.~\ref{fig_hg} below).

Photon production from a QGP at finite net quark density ($\mu_q$$>$0)
has been investigated for the tree-level diagrams in the upper left 
panel in Fig.~\ref{fig_qgplo} including resummed parton propagators 
in Ref.~\cite{Trax95}. To a good approximation, the net
effect can be represented by replacing the factor $T^2$ in
Eq.~(\ref{rate_pert}) by $(T^2+\mu_q^2/\pi^2)$.
Even under conditions for heavy-ion collisions at the SPS, where 
QGP formation at relatively high baryon density is conceivable, 
with $(\mu_q,T)\simeq(100,200)$~MeV, the $\mu_q^2$ term amounts
to a rather moderate correction (less than 5\%). 

In the not too far future one hopes to obtain nonperturbative 
information on photon production from a thermal QCD medium from
first principles via QCD lattice calculations. At the moment, these
calculations are available for the e.m. correlation function at
finite invariant mass $M$\cite{KL03}, \ie, for dilepton 
production.  In the limit of small masses one presently finds a 
less singular behavior than expected from perturbative 
calculations employing hard-thermal-loop resummation 
techniques\cite{BPY90}, implying a significant suppression of 
photon production rates. 
More definite conclusions have to await the use of larger lattices
to improve infrared sensitivities of the simulations. 

%%%%%%%%%%%%%%%%%%%%%%%%%%%%%%%%%%%%%%%%%%%%
\subsection{Hadronic Matter}
\label{ssec_hg}
%%%%%%%%%%%%%%%%%%%%%%%%%%%%%%%%%%%%%%%%%%%%
\begin{figure}[th]
\vspace{-0.8cm}
\hspace{-0.5cm}
\begin{minipage}{5.5cm}
\psfig{file=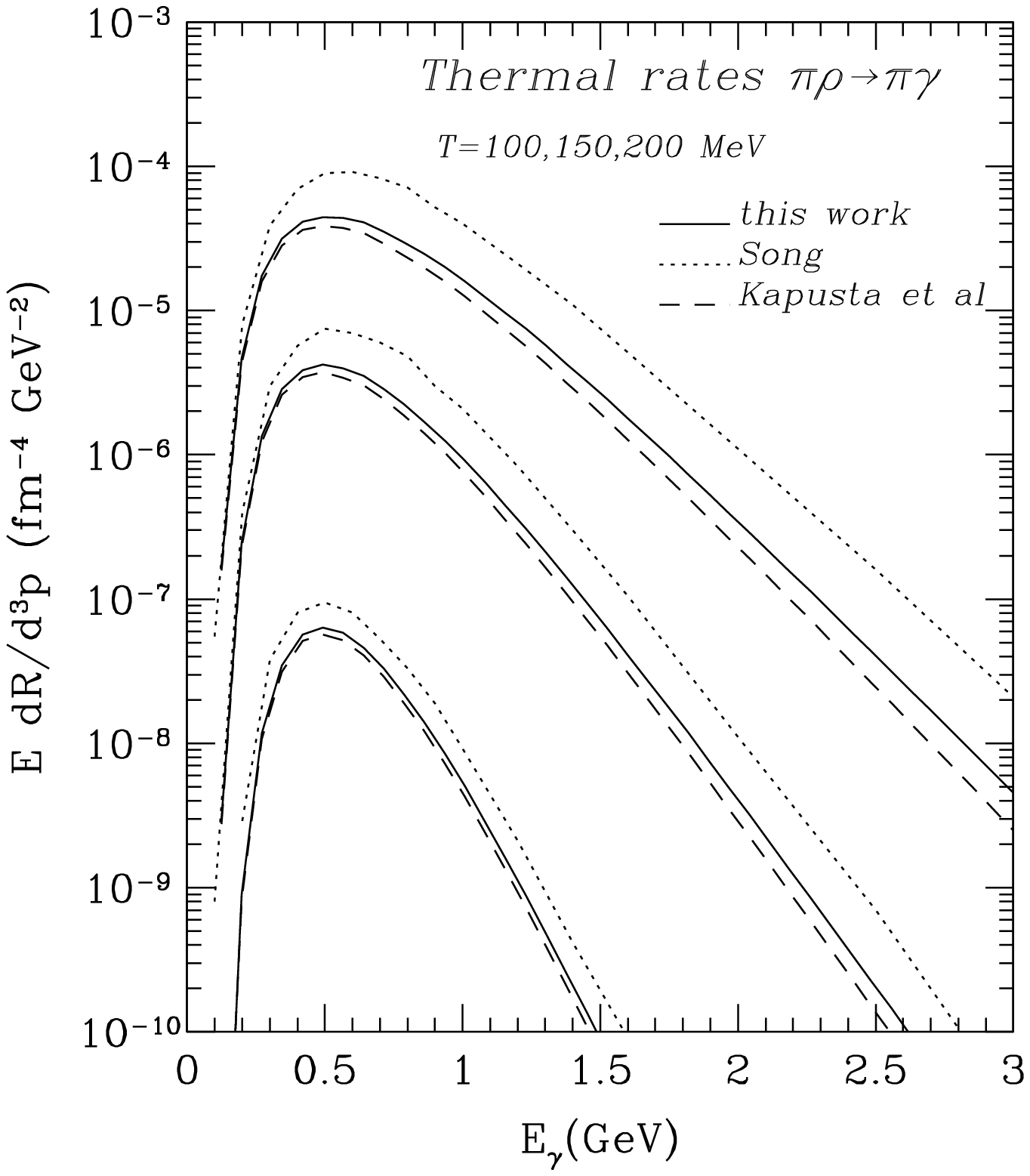,width=7cm,angle=0}
\end{minipage}
\begin{minipage}{6cm}
\psfig{file=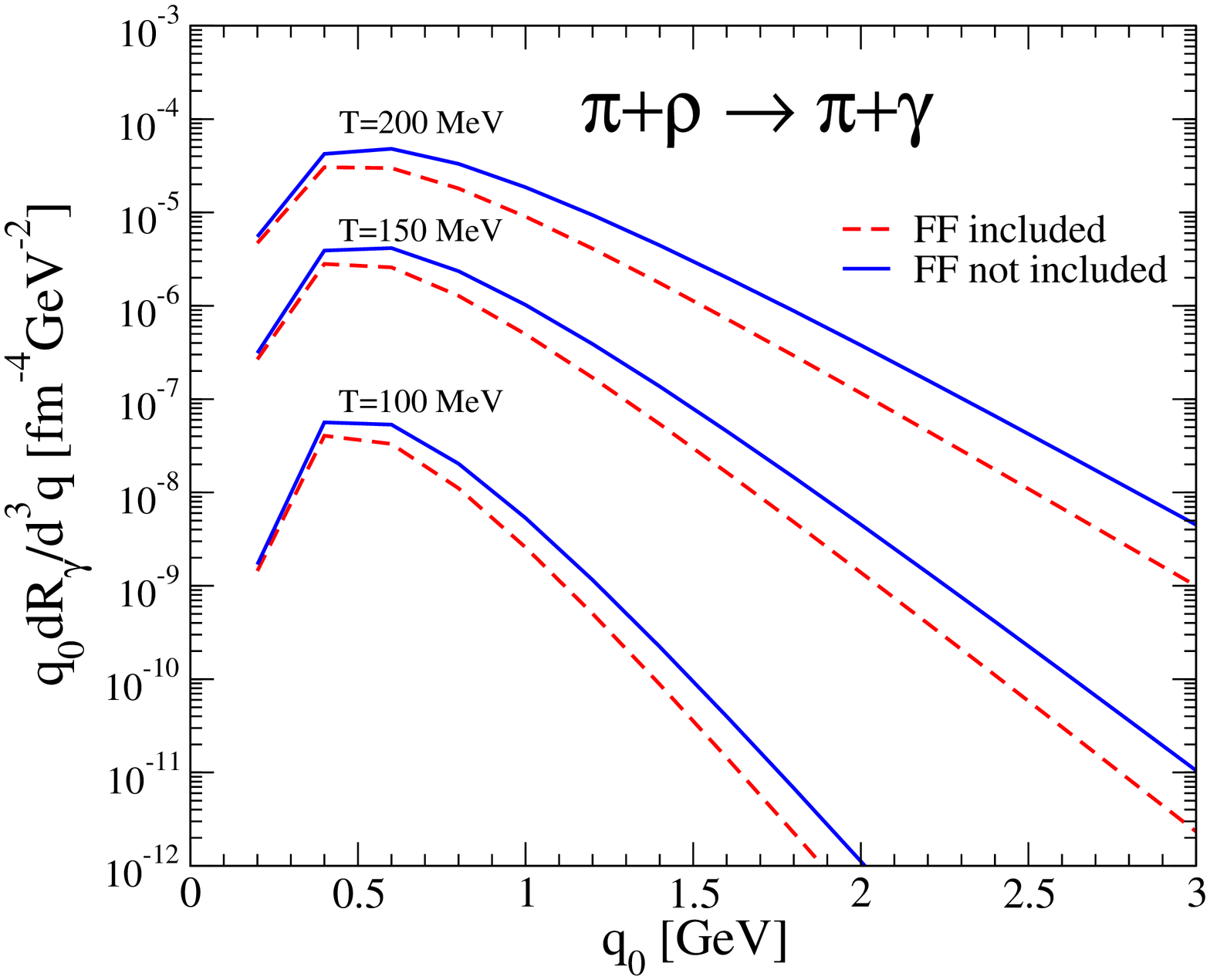,width=6.5cm,angle=-90}
\end{minipage}
\vspace*{8pt}
\vspace{-0.5cm}
\caption{Thermal photon production from inclusive
$\pi\rho\to\pi\gamma$ processes at the 2-loop level;
left panel: comparison of the rates (as taken from
Ref.~\protect\cite{Hal98}) obtained from a $\pi$-$\rho$
gas\protect\cite{KLS91} (dashed lines) and $\pi\rho a_1$ gas
within either the MYM\protect\cite{Song93} (dotted lines)
or the HLS framework\protect\cite{Hal98} (solid lines); all
calculations are without hadronic formfactors;
right panel: $\pi\rho a_1$ system within HLS\protect\cite{TRG04}
with (dashed lines) and without (solid lines) inclusion of hadronic
vertex formfactors.}
\label{fig_pirho}
\end{figure}
The study of thermal photon radiation from a hadronic gas is usually
carried out within effective Lagrangians. Constraints
on the interaction vertices can, to a certain extent, be imposed by 
symmetry principles (\eg, e.m. gauge and chiral invariance), and coupling 
constants are estimated by adjustment to measured decay branchings in 
the vacuum. Thus, for the temperature ranges 
relevant to practical applications, $T$=100-200~MeV, the predicted 
emission rates are inevitably beset with significant uncertainties, 
and therefore a careful judgment of the latter becomes mandatory. 

Investigations along these lines were initiated in Ref.~\cite{KLS91}, 
where the photon selfenergy has been computed to 
2-loop order for a mesonic system consisting of sharp (zero width) 
$\pi$-, $\eta$- and $\rho$-mesons (plus direct $\omega\to\pi^0\gamma$ 
decays). The important processes were identified as 
$\pi\pi\to\rho\gamma$ at low energy ($q_0$$\lsim$0.5~GeV) and  
$\pi\pi\to\rho\gamma$ at energies above $\sim$0.8~GeV (both proceeding 
via one-pion-exchange and a contact graph), as well as $\omega$ decays
around $q_0$$\simeq$0.5~GeV. It was also noted that effects of hadronic 
vertex formfactors could significantly affect the emission
rates especially at high energies. 
%(factors 2-3 suppression for $q_0$$\ge$1~GeV). 
In Ref.~\cite{XSB92} it was pointed out that 
$\pi\rho\to\pi\gamma$ scattering via $a_1(1260)$ resonance formation
(or, equivalently, $a_1\to\pi\gamma$ decay) constitutes an important 
contribution. This was followed up by a systematic treatment\cite{Song93}
 of an interacting $\pi\rho a_1$ system to 2-loop order within the 
Massive Yang-Mills (MYM) framework of introducing axial-/vector mesons
into a chiral Lagrangian, and, later,
within the Hidden-Local Symmetry (HLS) approach\cite{Hal98}.
A compilation of rates from the $\pi\rho\to\pi\gamma$
channel is shown in the left panel of Fig.~\ref{fig_pirho}.
The larger results of the MYM calculation as compared to the HLS one 
can be traced back to a significantly larger $a_1 \to \pi\gamma$ 
decay branching (which, in fact, overestimates the
empirical value). Indeed, a recent MYM calculation\cite{TRG04}
with a smaller $\Gamma_{a_1\to\pi\gamma}$ is consistent with
the previous HLS one, cf. right panel of Fig.~\ref{fig_pirho}.
\begin{figure}[th]
\centerline{
\psfig{file=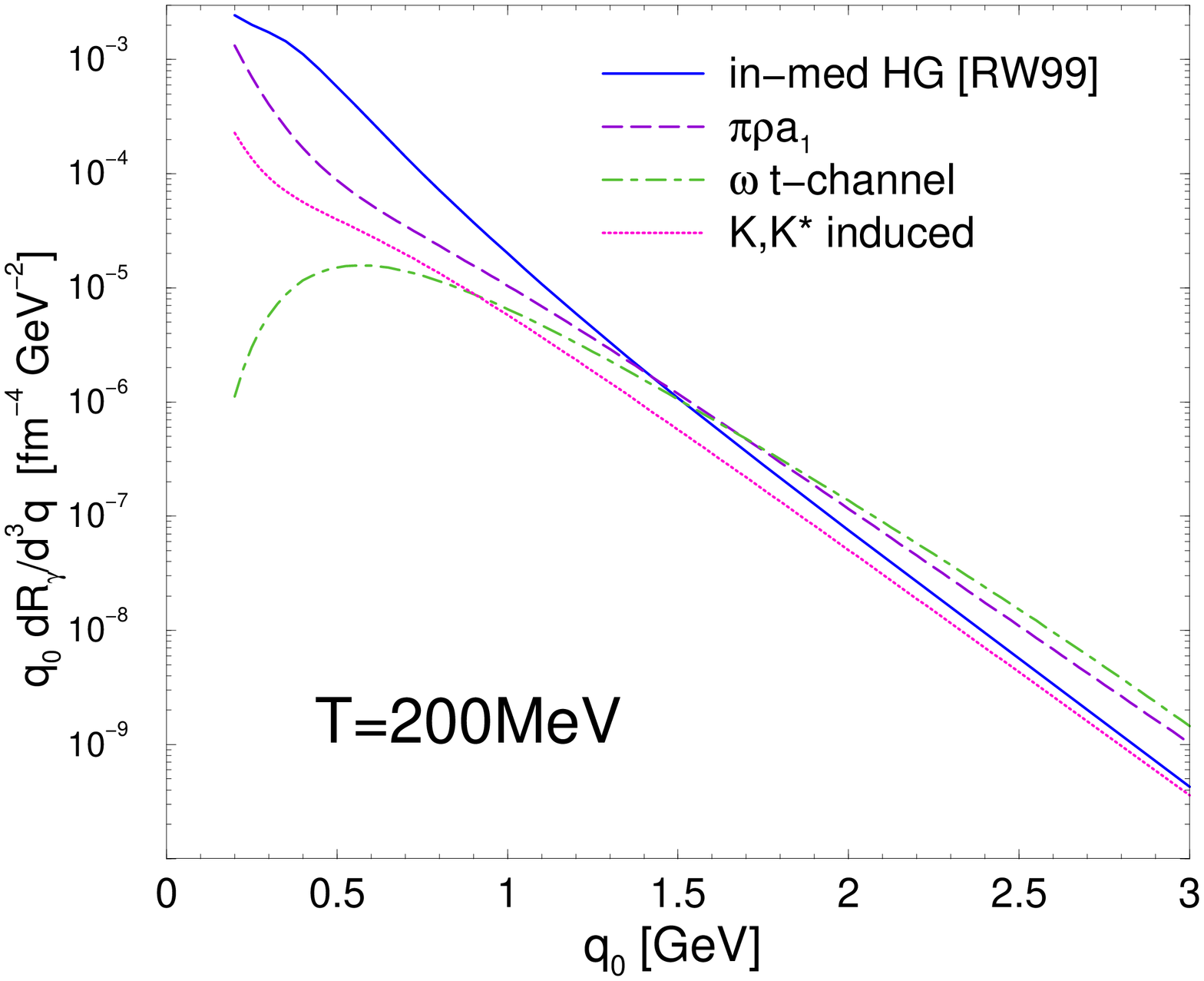,width=5.3cm,angle=-90}
\psfig{file=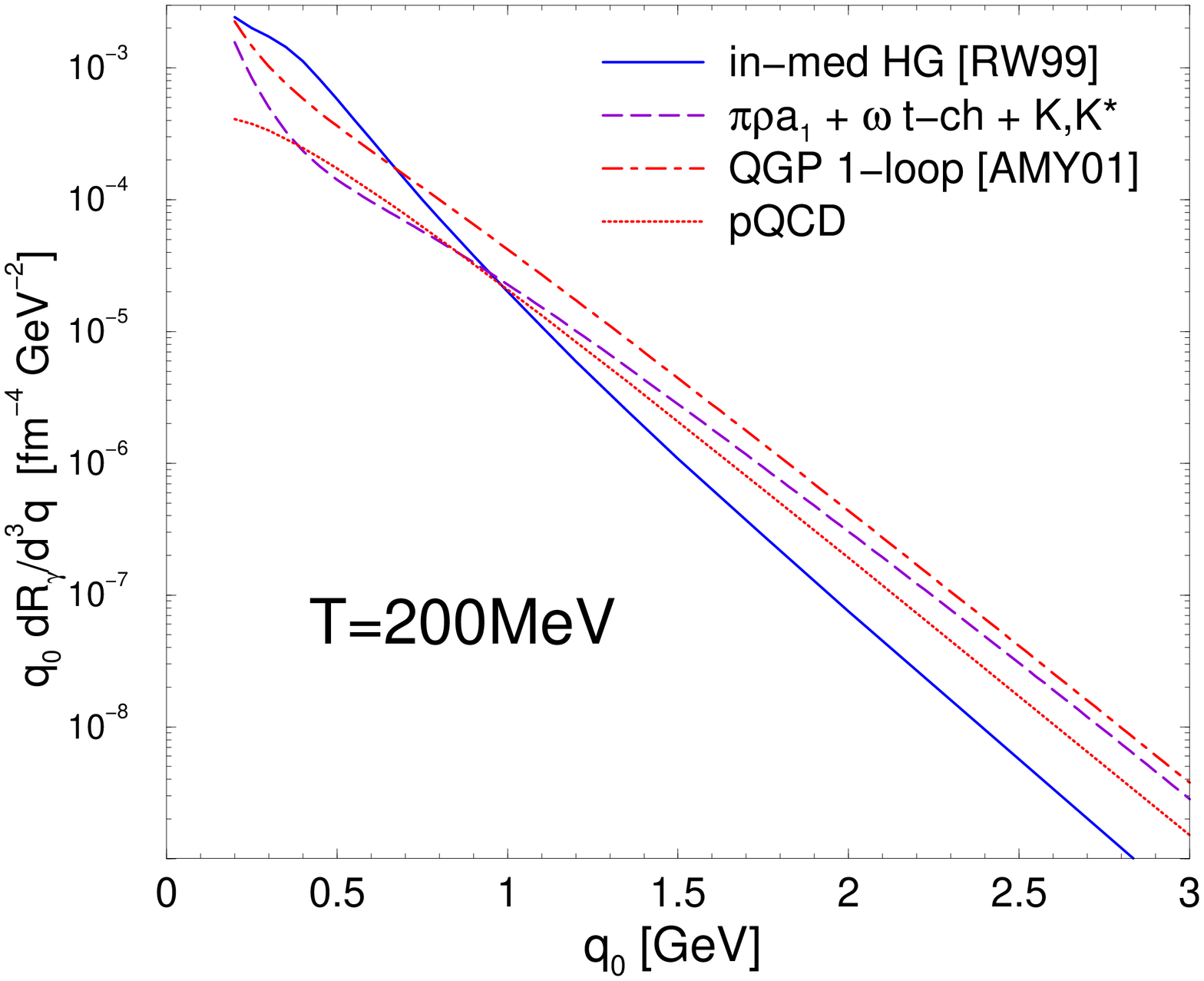,width=5.3cm,angle=-90}
}
\vspace*{8pt}
\caption{Left panel: thermal photon production channels
in hot and dense hadronic matter\protect\cite{TRG04} under 
conditions resembling URHICs at SPS energy;
right panel: total hadronic emission rates (sum of "in-med HG"
[solid line] and perturbative mesonic channels [dashed line])
compared to QGP ones (either the Born rate,
Eq.~(\ref{rate_pert}), or the complete leading order one,
Eq.~(\ref{rate_lo})).
}
\label{fig_hg}
\end{figure}
One should remark, however, that a consistent and accurate
vacuum phenomenology of the $\pi\rho a_1$ system (including radiative
and hadronic $a_1$ decays, as well as the $D/S$ ratio for the latter) has
not been achieved yet.  Also illustrated in the right panel of 
Fig.~\ref{fig_pirho} is the impact of hadronic formfactors\cite{TRG04}. 
As anticipated, a substantial reduction of the rate is found, reaching 
a factor 3-4 in the 2-3~GeV range, which should be included in
quantitative analyses.

Photon production from strangeness bearing mesons and baryonic matter
has been studied in Ref.~\cite{TRG04}. Within a HLS approach extended 
to flavor-$SU(3)$, $K$- and $K^*$-induced reactions contribute at the 
40-50\% level of the $\pi\rho a_1$ gas\cite{TRG04}, cf. left panel of
Fig.~\ref{fig_hg} (see also Ref.~\cite{Hag04}). 
Baryonic processes have been extracted from an 
in-medium $\rho$-meson spectral function\cite{RW99} via  
Eq.~(\ref{corr}), being quantitatively constrained by 
photoabsorption cross sections on nuclei\cite{RUBW98}. Under
conditions resembling $A$-$A$ collisions at the SPS, they 
constitute the dominant source of photons below $q_0$$\simeq$1~GeV.
Similar conclusions were reached in the chiral reduction formula 
framework\cite{SZ99}, whereas in Ref.~\cite{Alam03} baryonic 
contributions were found to be negligible for net-baryon free matter
(note that the photon selfenergy depends on the {\em sum} of
baryon and antibaryon densities which are not very different
at SPS and RHIC energies).
A surprisingly large photon yield has been identified in
Ref.~\cite{TRG04} as being due to $\omega$ $t$-channel exchange in
$\pi\rho\to\pi\gamma$, rendering it the single most important process
for energies beyond 2~GeV, after proper incorporation of hadronic
formfactors (dash-dotted line in the left panel of 
Fig.~\ref{fig_hg}). This result has been challenged in a 
recent comment\cite{Alam03b}, where, however, no hadronic 
formfactors were accounted for. 

The impact of in-medium modified hadron properties has been
studied in Refs.~\cite{SF98,ARSS03} in terms of reduced 
axial-/vector meson masses\cite{BR96}, leading to an increase 
of photon rates by up to a factor $\sim$10 close to $T_c$. In 
Ref.~\cite{Hal98} it has been argued within the HLS approach 
that in-medium coupling constants should also decrease, which
resulted in a small net change compared to the vacuum rates. 
Many-body effects on in-medium vector spectral functions are 
automatically incorporated when carrying the correlator to the 
photon point (solid lines in Fig.~\ref{fig_hg}).  

Finally, we compare in the right panel of Fig.~\ref{fig_hg} QGP with
HG emission: in the expected phase transition region, the top-down 
extrapolated partonic rates (especially for the complete LO case) are 
very close to the bottom-up extrapolated hadronic ones, quite reminiscent
to earlier results for thermal dilepton rates\cite{RW00}. 

%nonperturbative QGP: continuous transition in terms of the 
%dominant (low-lying) resonance contributions?)

%%%%%%%%%%%%%%%%%%%%%%%%%%%%%%%%%%%%%%%%%%%%%%%%%%%%%%%%%
\section{Current Status of Heavy-Ion Phenomenology}
\label{sec_hic}
%%%%%%%%%%%%%%%%%%%%%%%%%%%%%%%%%%%%%%%%%%%%%%%%%%%%%%%%%
To evaluate the role of thermal photon radiation in high-energy
heavy-ion collisions, an assessment of competing sources in the
experimental spectra is required. Since final-state 
hadron decays (most notably $\pi^0,\eta\to\gamma\gamma$) are
routinely removed from the measured spectra, the remaining 
nonthermal sources are essentially due to primary $N$-$N$ 
collisions and subsequent early evolution phases during 
which the matter has not yet thermalized. We will briefly discuss 
these sources in the following Section before turning
to the discussion of currently available data.

%%%%%%%%%%%%%%%%%%%%%%%%%%%%%%%%%%%%%%%%%%%%%%%%%%%%%%%%%%%%%%%%%%%%%%%%
\subsection{Non-Thermal Sources}
\label{ssec_nontherm}
%%%%%%%%%%%%%%%%%%%%%%%%%%%%%%%%%%%%%%%%%%%%%%%%%%%%%%%%%%%%%%%%%%%%%%%%
Photons produced in hard (primary) $N$-$N$ collisions (so-called 
"prompt" photons) are calculable from perturbative QCD
at sufficiently large transverse momentum (where they are
also expected to be most relevant, due to their 
power-law type spectra). Invoking the QCD factorization 
theorem, the photon-producing processes are in principle the same as 
in the QGP (recall left panel of Fig.~\ref{fig_qgplo}), but now 
convoluted over the parton distribution functions (PDF's) of the 
incoming hadrons. An additional component is due to
Bremsstrahlung processes in the fragmentation of a hard scattered
parton, where the corresponding fragmentation function is not under 
good control yet\cite{BFG98,cyr03}. Nevertheless, a satisfactory 
description of prompt photon data in $p$-$p$ collisions is possible, 
without having to introduce an intrinsic $k_t$ broadening in the 
nucleon PDF's. In $p$-$A$ collisions, additional nuclear 
effects occur, most notably the Cronin effect (also known as nuclear 
$k_t$-broadening) and shadowing of the PDF's. Whereas the former is 
most relevant (and experimentally well established) at fixed target 
energies ($\sqrt{s_{NN}}$$\le$50~GeV), the latter is likely to become 
increasingly important at high energies and forward rapidities, \ie, 
low at $x$ (as assessed, \eg, in recent calculations employing gluon 
saturation ideas\cite{GJ02,BMS04}). Indeed, at SPS energies 
($\sqrt{s}$$\le$ 20~GeV) the Cronin effect for prompt photons has been 
estimated to be rather significant\cite{Papp99,Dumi02}, whereas for 
central rapidities at RHIC neither nuclear broadening nor shadowing 
appear to be particularly prominent.  
    
The second nonthermal source in the early phases of URHICs is 
generically called the "pre-equilibrium" contribution which is 
theoretically rather difficult to evaluate. In both hadronic and 
partonic frameworks one key quantity is the formation time, either for 
building up hadronic wave functions, or for the thermalization of the 
QGP. The uncertainty of pre-thermal yields also 
pertains to their dependence on (transverse) momentum, 
as one does not expect (hadronic or partonic) modes to thermalize 
beyond a certain regime in $p_t$ (as borne out of hadronic 
spectra at RHIC (see, \eg, Ref.~\cite{qm04}), especially their 
azimuthal asymmetries).
One way to study the pre-equilibrium phase is by using parton cascade 
(or transport) simulations in connection with pQCD cross section (and 
quark fragmentation) for photon production\cite{SG98,Bass03}. The 
main problem in this framework is that many-body effects, \eg,  
quantum interference (LPM suppression) or color screening 
(required for infrared regularization), are not easily 
implemented in a controlled way. This can, in principle, be improved
within a quantum-field theoretical real-time formulation of an evolving 
partonic phase, as has been pursued in Ref.~\cite{Boya03}. Here, one
of the challenges is a realistic description of the bulk space-time
dynamics of the expanding system, starting from the incoming nuclei. 

Another conceivable nonthermal contribution can arise
from hard partons propagating through the QGP\cite{Fries03}, 
via either jet-photon conversion (gluon Compton scattering or
anti-/quark annihilation) or Bremsstrahlung off a quark. 
Generally speaking, all three pre-/off-equilibrium contributions
mentioned above lead to quite appreciable photon yields, 
as compared to thermal estimates, at momenta relevant for QGP 
radiation (cf. the following section).

%%%%%%%%%%%%%%%%%%%%%%%%%%%%%%%%%%%%%%%%%%%%%%%%%%%%%%%%
\subsection{Comparison to Data and Interpretations}
\label{ssec_comp}
%%%%%%%%%%%%%%%%%%%%%%%%%%%%%%%%%%%%%%%%%%%%%%%%%%%%%%%%
With photons being emitted throughout the evolution of a 
heavy-ion collision, a realistic description of the latter
is an important ingredient for determining the thermal yield,
especially if one studies sensitivities to the early (QGP)
phases.  One possibility are transport calculations
evaluating photon producing reactions on a process 
by process basis. The implementation of thermal production rates
as discussed in Sect.~\ref{sec_rates} is, however, most 
appropriately done in the context of hydrodynamic simulations, 
which are formulated in the same variables as the rates 
(temperature and chemical potentials for conserved charges). 
The main inputs for relativistic hydrodynamics are 
(i) the total initial energy and its spatial profile (according 
to, \eg, participant or primary-collision densities in the 
transverse plane), 
(ii) the initial 3-volume usually specified by the thermalization 
time $\tau_0$ as $V_0=\tau_0\Delta y \pi R_t^2$ ($R_t$: transverse
radius of the nuclear overlap for a given collision centrality), 
and, (iii) the equation of state governing the subsequent expansion, 
ideally taken from first principle lattice QCD (or in terms of 
suitable approximations thereof, \eg, quasiparticle QGP plus 
resonance hadron gas, matched via a latent heat at the transition). 
With the total initial energy being largely constrained by the
finally observed hadron multiplicities (or transverse energy), 
the key parameter in determining the QGP contribution is the
thermalization time $\tau_0$, which directly converts into an
initial temperature (note, however, that non-boost-invariant 
initial conditions in the longitudinal coordinate induce significant 
uncertainties in this relation, cf. Ref.~\cite{Huo02}).    
\begin{figure}[th]
\begin{minipage}{5.5cm}
\vspace{0.3cm}
\psfig{file=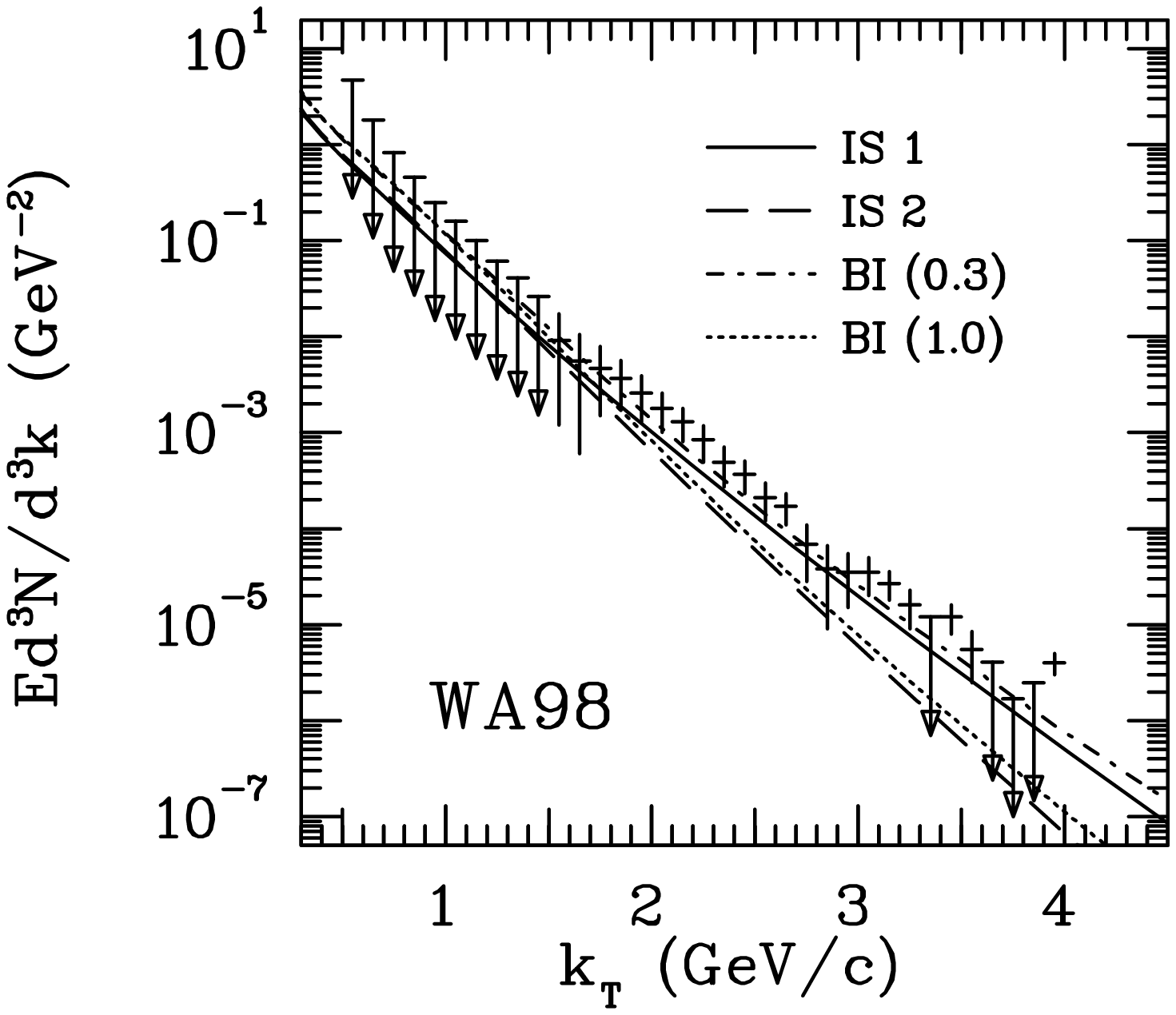,width=6cm}
\end{minipage}
\hspace{0.8cm}
\begin{minipage}{5.5cm}
\psfig{file=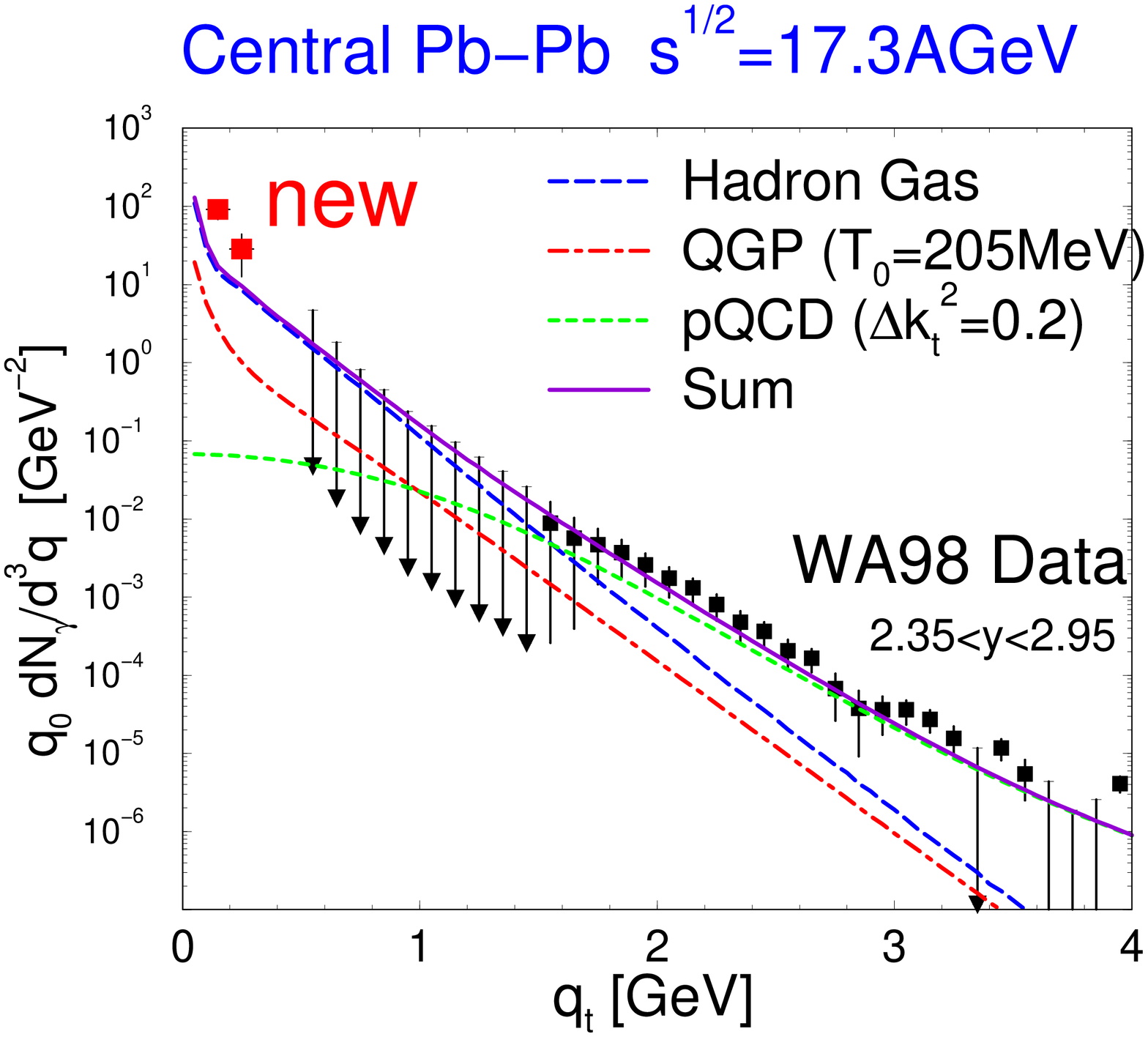,width=5.3cm,angle=-90}
\end{minipage}
\vspace*{8pt}
\caption{Direct photon spectra as measured by  
WA98\protect\cite{wa98-00,wa98-03} in central
$Pb$(158~AGeV)+$Pb$ collisions at the SPS.
Left panel: hydrodynamic calculations of thermal photon
spectra\protect\cite{Huo02} (including both QGP and HG emission)
assuming various initial conditions (dotted and dash-dotted line: 
boost invariant for $\tau_0$=1 and 0.3~fm/c, respectively; long-dashed 
and solid line: non-boost invariant longitudinal flow distributions
with average initial temperatures of $\bar T_0$$\simeq$215 and
255~MeV, respectively);
right panel: expanding thermal fireball calculations with
average initial temperature $\bar T_0$$\simeq$205~MeV,
supplemented with an initial pQCD yield that includes nuclear
$k_t$-broadening\protect\cite{TRG04}. }
\label{fig_sps}
\end{figure}
The hydrodynamic equations are solved on a space-time 
grid, and the thermal photon yield is straightforwardly obtained 
by multiplying the emission rate with the space-time (eigen-) volume 
of each unit cell at given local temperature and density.

An early measurement of photons in URHICs was performed by 
the WA80 collaboration 
in $S$(200AGeV)-$Au$ collisions at the SPS, resulting in
upper limits on the direct photon yield\cite{wa80}. 
The theoretical consensus was that in these reactions
initial temperatures in excess of 250~MeV could be 
excluded\cite{Dumi95,SS00}.
The successor experiment WA98 succeeded in obtaining a nonzero 
signal resulting in the direct-photon $q_t$-spectra\cite{wa98-00} 
as shown in Fig.~\ref{fig_sps}. 
In the left panel, the data are compared to hydrodynamic 
calculations\cite{Huo02} which include both QGP and hadronic phase 
with a transition at $T_c$=165~MeV, employing complete leading order 
emission rates  in the QGP\cite{AMY01} and a HG rate based on 
$\pi\pi\to\rho\gamma$ and $\pi\rho\to\pi\gamma$ according to 
Refs.~\cite{KLS91,XSB92} (including $\pi$ and $s$-channel $a_1$ 
exchanges).  
The comparison of different initial states (non-/boost invariant, 
with formation times of 0.3 and 1.0~fm/c) indicates that
a satisfactory description of the WA98 in terms of thermal radiation
(without contributions from primary $N$-$N$ collisions) 
requires rather large initial temperatures $T_0^{max}$$\simeq$330~MeV
(translating into $\bar{T}_0$$\simeq$260~MeV when averaged over the 
transverse plane), rendering the QGP radiation the dominant source
(similar results have been obtained in Ref.~\cite{SS01}). 
These conclusions are not significantly altered if the primordial
pQCD yield is added in terms of $N$-$N$ collision-scaled spectra
from $p$-$p$ collisions. The situation changes if the latter
contribution is augmented by accounting for the Cronin effect,
cf. right panel of Fig.~\ref{fig_sps}. With a rather moderate
nuclear $k_t$-broadening, the initial pQCD yield increases by about
a factor of 2-3 in the $q_t$=2-4~GeV range, and thus turns out to 
saturate the experimental spectra for $q_t$$>$2.5~GeV. Consequently, 
the role of the thermal yield, which has been modeled in a more
simplistic thermal fireball evolution, is diminished. With the
same underlying QGP emission rates as in the hydrodynamic calculation 
(but a more complete description of the hadronic rates\cite{TRG04}) 
the data are compatible with significantly smaller (average)
initial temperatures of $\bar T_0$$\simeq$205~MeV (corresponding to
$\tau_0$$\simeq$1~fm/c), rendering the QGP contribution subdominant
in the relevant range $q_t$$\le$2.5~GeV. In fact, this can also 
be expected to hold for the hydrodynamic calculations.   
  
A rather dramatic addition to the WA98 data has recently been provided
at low transverse momenta ($q_t$=100-300~MeV, labeled "new" in
the right panel of Fig.~\ref{fig_sps}), extracted via $\gamma$-$\gamma$
HBT interferometry techniques\cite{wa98-03} (which automatically 
eliminate late hadron decays, most notably $\pi^0,\eta\to\gamma\gamma$).
The default thermal calculations underestimate these data by a rather
substantial margin. Some improvement arises when including
soft $\pi\pi\to\pi\pi\gamma$ Bremsstrahlung, via $S$-wave $\pi\pi$
interactions\cite{TGR04} (which are suppressed at higher $q_t$). 
Additional
hadronic processes with large yields are not easily conceivable.
Increasing the fireball lifetime by, \eg, 30\%, implying lower
thermal freezeout temperatures $T_{fo}$$\simeq$90~MeV (to
be compared to standard values of $\sim$~110MeV), 
%(which maybe arguable for very soft interactions), 
increases the thermal contribution at low $q_t$ by about the same
fraction.  This raises the exciting possibility that one is observing 
substantial in-medium modifications, \eg, a reduced "$\sigma$"-mass
in the $S$-wave $\pi\pi$ interaction. A possibly related phenomenon 
may have been observed in $\pi$- and $\gamma$-induced 2$\pi$ 
production experiments off ground state 
nuclei\cite{chaos96,cryst00,taps02,Kuni04}.  
The measured invariant-mass distributions in the scalar-isoscalar 
channel exhibit appreciable shifts of strength towards the 2-pion 
threshold when going from hydrogen to heavy nuclear targets (no such 
effect is observed in the pure isotensor $\pi^+\pi^+$ final-state 
where the (strong) $\pi\pi$ interaction is repulsive), 
despite expected pion absorption (which biases pion emission towards 
the nuclear surface, \ie, rather small densities).
A similar softening in the $\pi^-\pi^+$ interaction in the heavy-ion
environment, where additional thermal occupation factors appear,
could induce a substantial soft-photon enhancement\cite{TGR04}.
Another possibility are medium modifications of the $\Delta$(1232)
resonance\cite{HR04}, which has a large radiative branching ratio. 
% comment on Low theorem

\begin{figure}[th]
\vspace{-0.3cm}
\centerline{
\hspace{0.1cm}   \psfig{file=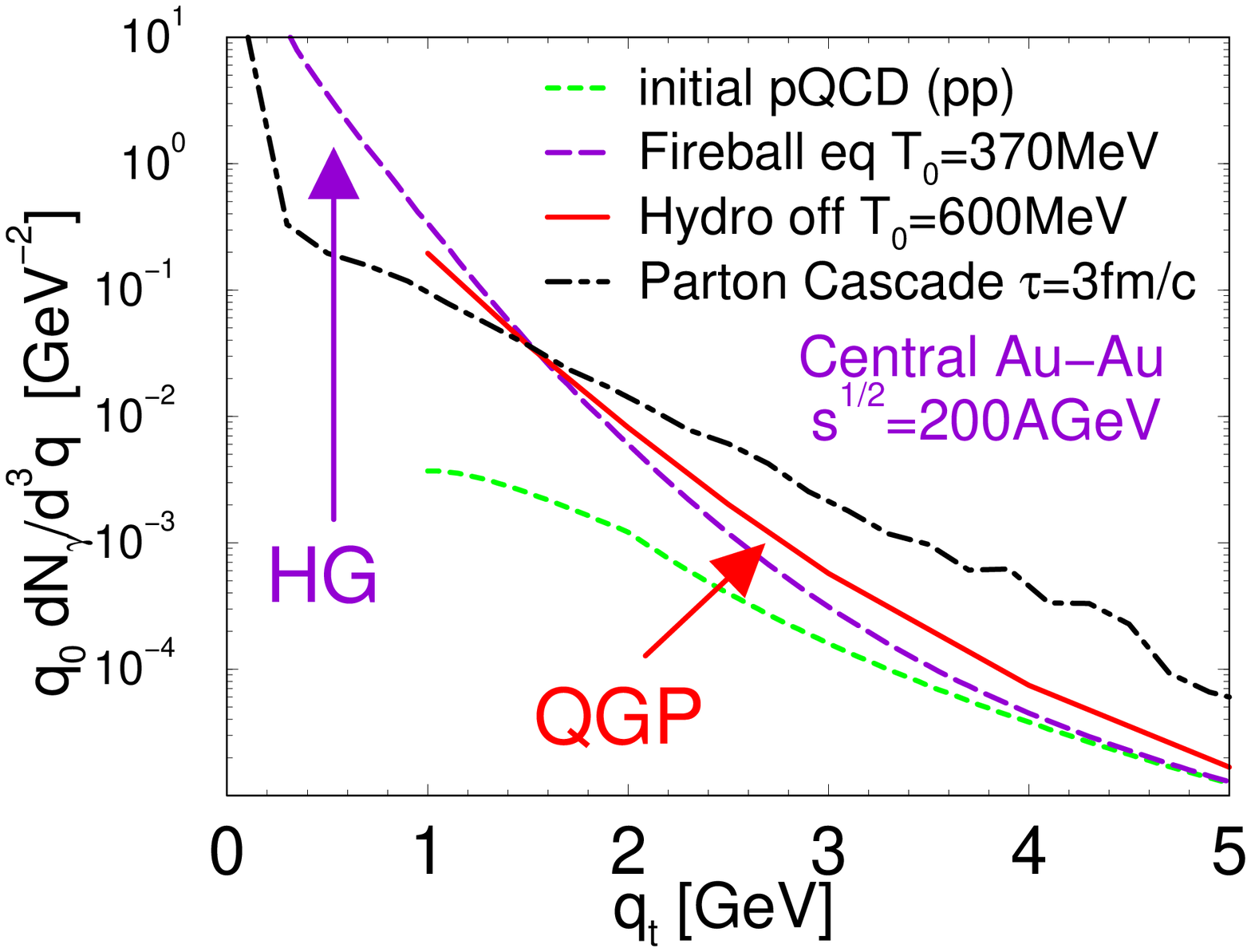,width=6.1cm}
\hspace{-0.3cm}  \psfig{file=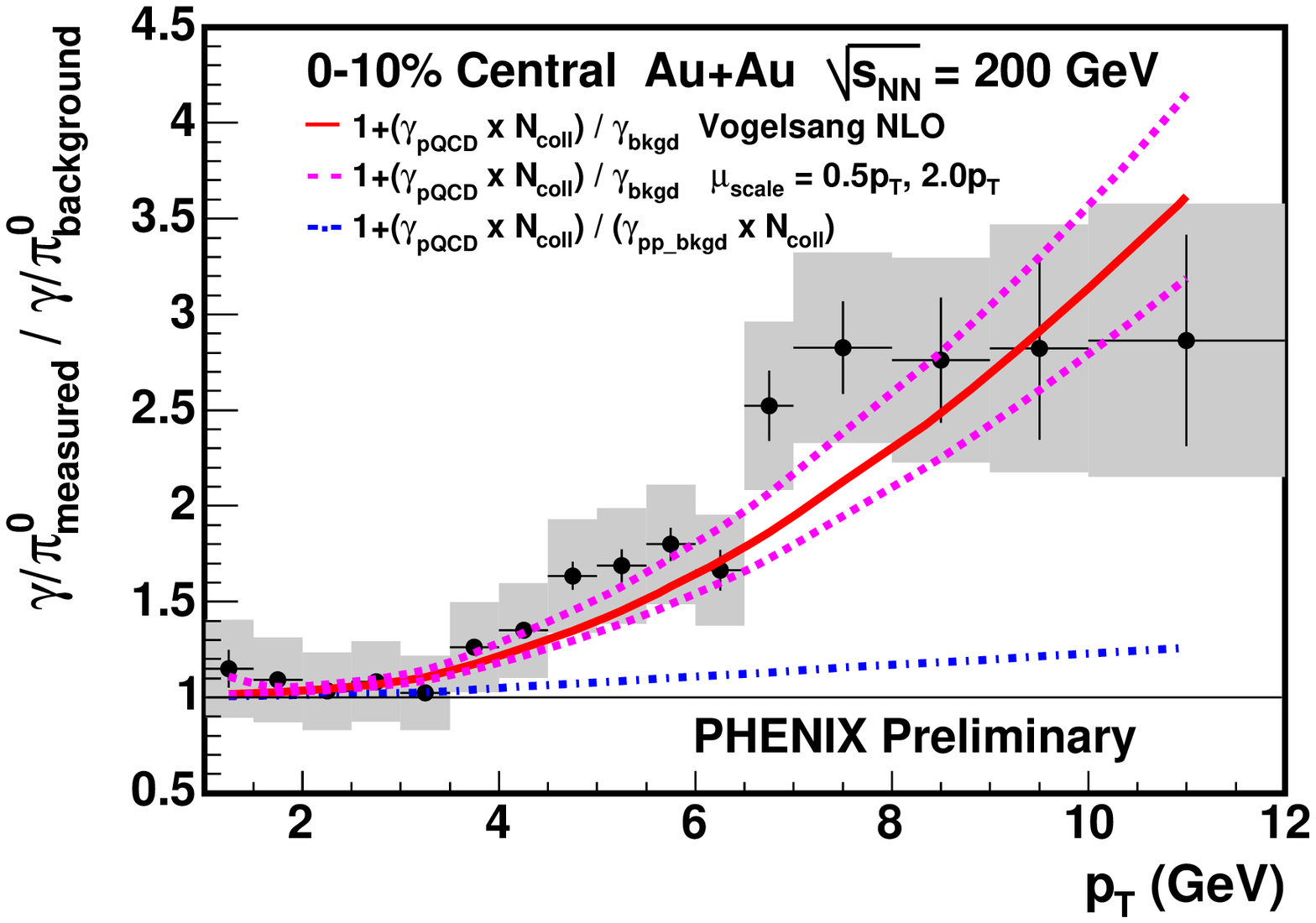,width=7.2cm}}
\vspace*{8pt}
\caption{Left panel: compilation\protect\cite{Ra04} of 
direct photon yields for central $Au$-$Au$ collisions at RHIC:
hydrodynamic\protect\cite{Niemi04,cyr03} (solid line) and 
fireball\protect\cite{TRG04} (long-dashed line) calculations
of thermal radiation including the yield from primordial pQCD
photons\protect\cite{Sriv01} (shown also separately by the 
short-dashed line), as well as results from the first 3~fm/c of 
a parton cascade simulation\protect\cite{Bass03} (dash-dotted line);
right panel: direct photon data\protect\cite{phenix04} from
central $Au$-$Au$ at RHIC compared
to next-to-leading order pQCD predictions\protect\cite{Vogel99}
for primordial $N$-$N$ collisions (scaled by the number of hard
collisions).
}
\label{fig_rhic}
\end{figure}
Direct photon spectra as predicted for central 
$Au$-$Au$ collisions at RHIC energy are compiled in the left panel 
of Fig.~\ref{fig_rhic}\cite{Ra04}. Thermal 
radiation from two different approaches, \ie, 
hydrodynamic\cite{cyr03,Niemi04} (solid line, for formation time 
$\tau_0$=0.18~fm/c) and expanding fireball\cite{TRG04} (long-dashed 
line, for formation time $\tau_0$=0.33~fm/c), is in approximate 
agreement with each other (the 2 curves also include primordial
pQCD contributions). In both calculations QGP radiation 
dominates for transverse momenta above $q_t\simeq$1~GeV. 
Whereas in the fireball expansion the QGP has been assumed
to be in chemical equilibrium, the hydrodynamic evolution 
incorporates an undersaturation of gluons and especially 
quarks (which is motivated by the predominance of low-$x$ 
gluons in the wave function of the incoming nucleons, and 
represented by initial fugacities $\lambda_{q,\bar q,g}$$<$1,
evolved in time) in both equation of state and QGP emission 
rates\cite{cyr03,Niemi04}. Compared to the chemical equilibrium 
case, parton undersaturation implies reduced emission rates, but 
also larger initial temperatures if the initial entropy, 
which determines the number of produced particles, is fixed. 
It turns out that, to a large
extent, these two effects compensate each other\cite{cyr03}.    

The primordial pQCD contribution in Fig.~\ref{fig_rhic} (left
panel) is taken from a parametrization\cite{Sriv01} of $p$-$p$ 
data over a large range of center-of-mass energies, scaled by
the number of $N$-$N$ collisions in central 
$Au$-$Au$, and without Cronin effect 
(expected to be rather small at RHIC). 
Comparison with the thermal contribution indicates that the latter 
can compete up to $q_t$$\simeq$5~GeV.
The dash-dotted line in Fig.~\ref{fig_rhic} (left panel) represents
an estimate of photon production from parton rescattering and 
fragmentation within the first 3~fm/c of a parton cascade 
model\cite{Bass03}. The resulting photon yield exceeds both thermal
and primordial pQCD yields by up to an order of magnitude
for transverse momenta of around 4-5~GeV. In this regime,  
the contribution from jet-$\gamma$ conversions\cite{Fries03}
(not shown in Fig.~\ref{fig_rhic}) turns out to be comparable in 
magnitude to the thermal+pQCD radiation.  

In the right panel of Fig.~\ref{fig_rhic} preliminary data from 
PHENIX\cite{phenix04} are displayed as a (double) ratio of measured 
photons over those expected from meson decays. 
Above $q_t$$\simeq$4~GeV,
this ratio significantly exceeds one, being attributed to direct 
photons, which are in good agreement with predictions  
for primordial pQCD photons\cite{Vogel99}. 
Currently there is not enough sensitivity to QGP radiation, 
but the large yields obtained in the parton cascade are disfavored.

The prospects of identifying thermal photon radiation in future 
heavy-ion collisions at the LHC ($Pb$-$Pb$ at $\sqrt{s}$=5500AGeV)
have been thoroughly assessed in a recent CERN Yellow 
Report\cite{cyr03}. These studies suggest that the window in which 
thermal photons exceed the pQCD production could extent up to almost 
$q_t$=10GeV, with larger maximal enhancement factors than at RHIC. 

%%%%%%%%%%%%%%%%%%%%%%%%%%%%%%%%%%%%%%%%%%%%%%%%%%%%%%%%%%%%
\section{High-Density Quark Matter}
\label{sec_csc}
%%%%%%%%%%%%%%%%%%%%%%%%%%%%%%%%%%%%%%%%%%%%%%%%%%%%%%%%%%%%
Besides high-energy heavy-ion collisions (and the early universe),
the other systems in nature in which strongly interacting
matter under extreme conditions is likely to exist are compact
stars. Contrary to the high-temperature medium formed
in URHICs, compact stars are characterized by high baryon
densities at {\em low} temperatures.  
Nevertheless, much like for URHICs, electroweak emission 
spectra play an important role in inferring properties of the 
matter within the star. The main focus is traditionally on 
neutrinos\cite{Prak01}, due to their substantially longer mean 
free path as compared to photons. 
Renewed interest in the electroweak emission spectra was 
generated when it was realized that color-superconducting quark 
matter (CSC) may exist with values for the quark-pairing
as large as $\Delta$$\sim$100~MeV\cite{RSSV98,ARW98}, rendering 
them possibly relevant for phenomenological applications. 
Subsequently, both neutrino\cite{Blasch00,Page00,JPS02,RST03} 
and photon\cite{Jai02,Vogt04} emissivities have been assessed 
for CSC. Clearly, if CSC is limited to exist in 
the core of compact stars, pertinent photon signals cannot 
be observed. The latter can therefore only be relevant for 
(hypothetical) 
"strange quark matter" stars, in which quark matter extents
all the way to the star's surface. Even in this case, not much 
is known as to how the surface structure impacts photon emission. 

This not withstanding, thermal photon rates from the so-called
color-flavor-locked (CFL) phase\cite{ARW99} of CSC  
were calculated in Refs.~\cite{Jai02,Vogt04}. 
In CFL matter, $u$, $d$ and $s$ quarks of approximately equal 
Fermi momenta form Cooper pairs in a way that breaks the original
$SU(3)_C$$\times$$SU(3)_L$$\times$$SU(3)_R$ symmetry to a global
$SU(3)$ one, implying (among other things) the emergence
of 8 Goldstone bosons. The small explicit breaking due
to the finite current quark masses induces corresponding pion 
masses $m_\pi$$\approx$10~MeV (even smaller (larger)
for (anti-) kaons). For temperatures (well) below this scale,
Goldstone boson excitations are exponentially suppressed and
photon emissivities are controlled by electromagnetic processes
involving $e^\pm$ and $\gamma$'s, provided an electrosphere forms
at the star surface\cite{Page02,Jai04} (this is, in fact, not
required for a pure CFL star, which is electrically neutral
without the presence of electrons).  
Here, we focus on situations where the strong-interaction degrees
of freedom are active, \ie, temperatures of order tens of MeV as 
characteristic for the early evolution phases following a supernova
explosion. 

The photon selfenergy in CSC has been computed in 
Ref.~\cite{LM01} at zero temperature and for asymptotic densities
employing weak coupling techniques. The corresponding $q\bar q$ 
one-loop selfenergy in the CFL phase has been extended to 
finite temperatures and applied to photon emission rates using 
Eq.~(\ref{corr}) in Ref.~\cite{Jai02}. For temperatures
$T$$\sim$70~MeV and comparable baryon densities, the 
rates are quite similar to extrapolations of hadronic many-body 
calculations\cite{RW99} (cf. Sect.~\ref{ssec_hg}).

\begin{figure}[th]
\centerline{
\psfig{file=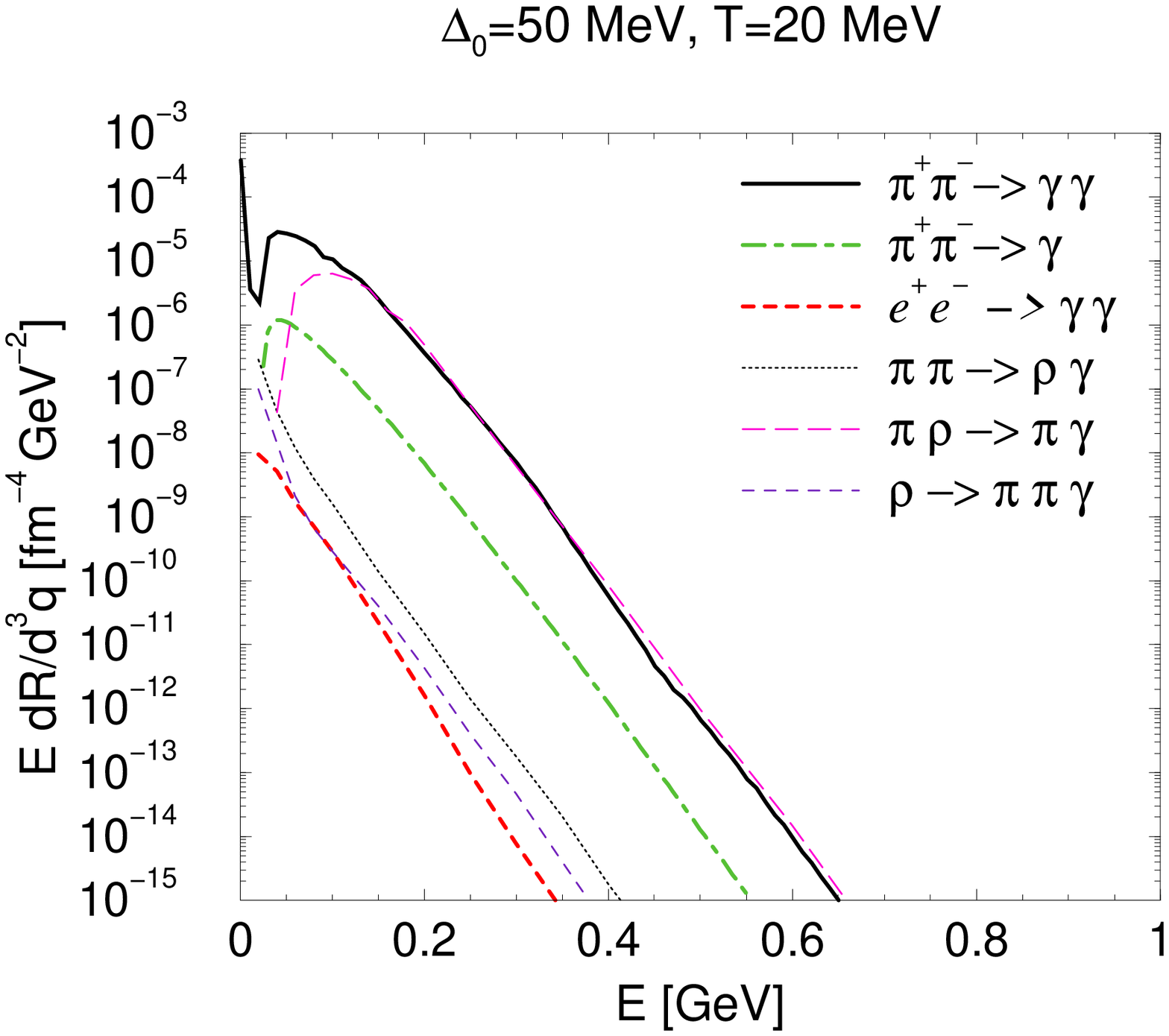,width=6cm,angle=0}
\hspace{0.3cm}
\psfig{file=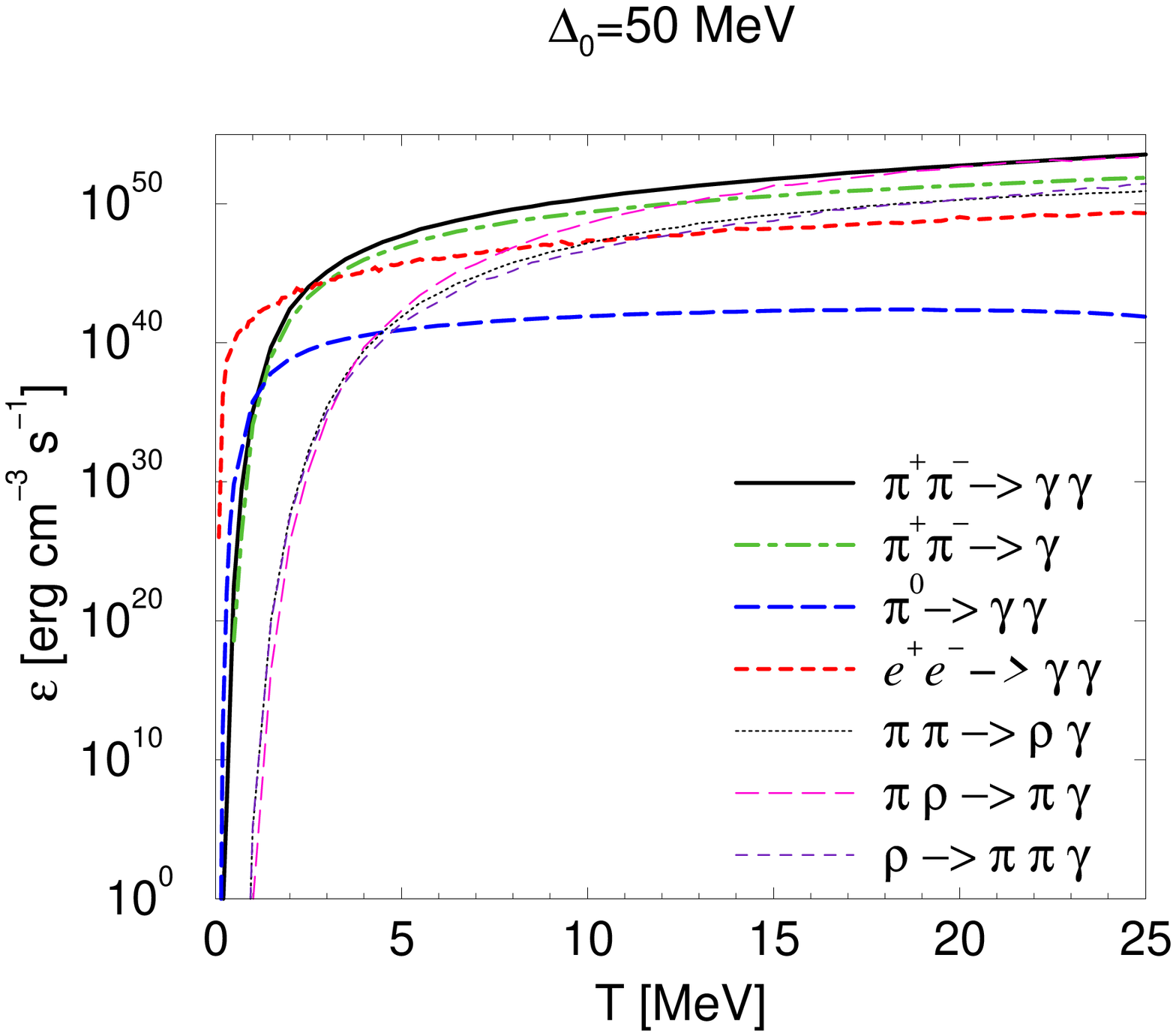,width=6cm,angle=0}
}
\vspace*{8pt}
\caption{Photon emission rates from CFL quark matter
within a HLS approach (left panel), and corresponding (integrated)
emissivities vs. temperature\protect\cite{Vogt04}.
The finite-$T$ gap value has been assumed to follow the BCS
relation $\Delta(T)$=$\Delta_0\sqrt{1-(T/T_c)^2}$
with $T_c$=0.57$\Delta_0$.
}
\label{fig_csc}
\end{figure}
In Ref.~\cite{Vogt04}, the notion of Goldstone bosons in the CFL 
phase has been exploited to formulate a HLS framework in the strong
coupling regime. In addition, an in-medium pion dispersion 
relation\cite{Son00}, $\omega^2=m_\pi^2+v_\pi^2p^2$ with 
$v_\pi^2=1/3$, was accounted for, leading to novel annihilation 
(and decay) processes of type $\pi^+\pi^-\leftrightarrow \gamma$ 
(see also Ref.~\cite{RST03}).  With a $\rho$-meson mass of 
$\sim$2$\Delta$\cite{RSWZ00}, and assuming VDM, the gauge coupling 
was found to be $\tilde{g}$=$g_{\rho\pi\pi}$$\simeq$2, 
significantly smaller than its vacuum value of $\sim$6.
The rates and their integrated emissivities in Fig.~\ref{fig_csc}
indicate that processes induced by Goldstone bosons dominate
over $e^+e^-$ annihilation at temperatures as low as 5~MeV; even
strong-interaction processes involving $\rho$ mesons may become
irrelevant for temperatures $T$$\simeq$20~MeV, and thus 
for newborn compact stars. In particular, the results for
photon mean-free-path and emissivity imply that the photon flux
from a (hypothetical) hot CFL star saturates the black-body limit.

%%%%%%%%%%%%%%%%%%%%%%%%%%%%%%%%%%%%%%%%%%%%%%
\section{Conclusions}
\label{sec_concl}
%%%%%%%%%%%%%%%%%%%%%%%%%%%%%%%%%%%%%%%%%%%%%
In recent years significant progress has been made in our
attempts to address the question: what is the electromagnetic
blackbody spectrum of strong-interaction matter at given
temperature and density?
In the QGP phase, the role of soft $t$-channel exchanges has
been understood better, leading to a photon emission rate that
constitutes the full result to leading order in $\alpha_s$. 
In the hadronic phase, existing calculations could be refined
and supplemented by additional processes towards a more
complete description.  
Clearly, in the vicinity of the expected (pseudo-) phase transition
neither current partonic nor hadronic calculations are quantitatively
reliable, but the fact that they agree within, say, a factor
of two over a large range of energies (over which the rate itself 
changes by several orders of magnitude), increases the confidence
in both calculations, and may eventually teach us something about
the phase transition itself. Fully nonperturbative results from 
(unquenched) QCD lattice calculations are eagerly awaited. 
Moreover, phenomenological applications to (ultra-) relativistic
heavy-ion collisions compare favorably to existing data (except
the new low-momentum data from WA98), further corroborating the
(approximate) validity of theoretical estimates. At the SPS, 
most approaches agree that temperatures in excess of the
critical one are required in the early phases of head-on 
$Pb$-$Pb$ collisions. This adds significantly to
similar evidence gained from dilepton spectra. At RHIC, the QGP 
signal is expected to become substantially stronger, but so do 
prompt photon yields from initial hard collisions.  
Up to now, electromagnetic signals from URHIC's have never failed
to generate excitement. 

\section*{Acknowledgments}
It is a pleasure to thank
C. Gale, P. Jaikumar, R. Ouyed, S. Turbide, C. Vogt and I. Zahed 
for enjoyable and stimulating collaboration on the presented topics.
Interesting discussions with T. Awes and K. Haglin 
are also acknowledged.

\end{document}